\documentclass[apj]{emulateapj}
\usepackage{apjfonts}
\usepackage{url}
\usepackage{multirow}
\usepackage{enumerate}
\usepackage{hyperref}

\usepackage{comment}
 \usepackage{color}           % For color text: \color command
 \definecolor{DarkGreen}{rgb}{0.0,0.45,0.0}  % define a custom color

\newcommand{\sat}[1]{\it\uppercase{#1}\rm}
\newcommand{\fig}[1]{Figure~\ref{#1}}

\newcommand{\speed}[1]{#1 km~s${}^{-1}$}

\begin{document}

\shorttitle{Double-Decker Filament I} %

\shortauthors{Liu et al.}

\title{Slow Rise and Partial Eruption of a Double-Decker Filament.\\ 
       I Observations and Interpretation}

\author{Rui Liu\altaffilmark{1,2}, Bernhard Kliem\altaffilmark{3,4}, Tibor T\"{o}r\"{o}k\altaffilmark{5},
Chang Liu\altaffilmark{2}, Viacheslav S. Titov\altaffilmark{5}, \\
Roberto Lionello\altaffilmark{5}, Jon A. Linker\altaffilmark{5}, and Haimin Wang\altaffilmark{2,6}}

\altaffiltext{1}{CAS Key Laboratory of Geospace Environment, Department of Geophysics \& Planetary Sciences, University of Science \&
Technology of China, Hefei 230026, China}

\altaffiltext{2}{Space Weather Research Laboratory, Center for Solar-Terrestrial Research, NJIT,
Newark, NJ 07102, USA}

\altaffiltext{3}{Institute of Physics and Astronomy, University of Potsdam, 14476 Potsdam,
Germany}

\altaffiltext{4}{Mullard Space Science Laboratory, University College London,
                 Holmbury St.~Mary, Dorking, Surrey RH5 6NT, UK}

\altaffiltext{5}{Predictive Science Inc., 9990 Mesa Rim Road, Suite 170, San Diego, CA 92121, USA}

\altaffiltext{6}{Key Laboratory of Solar Activity, National Astronomical Observatories, Beijing, China}

\email{rliu@ustc.edu.cn}

%\slugcomment{Manuscript kink\_part1a; in prep.\ for ApJ}
%\date{\today}
\journalinfo{accepted for publication in ApJ}
%\submitted{Draft version: \today}

\begin{abstract}
We study an active-region dextral filament which was composed of two branches separated in height by about 13 Mm, as inferred from three-dimensional reconstruction by combining \sat{sdo} and \sat{stereo-b} observations. This ``double-decker'' configuration sustained for days before the upper branch erupted with a \sat{goes}-class M1.0 flare on 2010 August 7. Analyzing this evolution, we obtain the following main results.
1) During hours before the eruption, filament
threads within the lower branch were observed to intermittently brighten up, lift upward, and then
merge with the upper branch. The merging process contributed magnetic flux and current to the
upper branch, resulting in its quasi-static ascent.
2) This transfer might serve as the key mechanism for the upper branch to lose
equilibrium by reaching the limiting flux that can be stably held down by the overlying field or by
reaching the threshold of the torus instability.
3) The erupting branch first straightened from a reverse S shape that followed the polarity
inversion line and then writhed into a forward S shape. This shows a transfer of left-handed
helicity in a sequence of writhe-twist-writhe. The fact that the initial writhe is converted
into the twist of the flux rope excludes the helical kink instability as
the trigger process of the eruption, but supports the occurrence of the instability in the main phase,
which is indeed indicated by the very strong writhing motion.
4) A hard X-ray sigmoid, likely of coronal origin, formed in the
gap between the two original filament branches in the impulsive phase of the associated flare.
This supports a model of transient sigmoids forming in the vertical flare current sheet.
5) Left-handed magnetic helicity is inferred for both branches of the dextral filament.
6) Two types of force-free magnetic configurations are compatible with the data, a double flux
rope equilibrium and a single flux rope situated above a loop arcade.

\end{abstract}


\section{Introduction}

It is generally accepted that the magnetic field plays a crucial role for dense and cold filaments
to be suspended in and thermally isolated from the surrounding hot, tenuous coronal plasma.
Filaments are always formed along a polarity inversion line (PIL) of the photospheric field.
Studies utilizing Zeeman and Hanle effects demonstrated that the flux threading the filament
is largely horizontal and mainly directed along the filament axis \citep{leroy89, bommier94}. This is also
manifested in the chromospheric fibril pattern \citep[][and references therein]{martin98}: fibrils
near the filament are nearly parallel to the filament axis, but away from the filament they tend to
be perpendicular to the filament axis. This pattern implies the presence of two types of
\emph{filament chirality}:
for an observer viewing the filament from the positive-polarity side, the axial field in a
\emph{dextral} (\emph{sinistral}) filament always points to the right (left). Independent of the
solar cycle, dextral (sinistral) filaments are predominant in the northern (southern) hemisphere
\citep{mbt94, Zirker&al1997, pbr03}.

Most quiescent filaments have what is known as \emph{inverse polarity} configuration, i.e., the
magnetic field component perpendicular to the axis traverses the filament from the region of
negative polarity to the region of positive polarity in the photosphere,
opposite to what would be expected from a potential field. \emph{Normal
polarity} configuration is mainly found in active-region filaments, i.e., the field lines pass
through the filament from the region of positive polarity to the region of negative polarity
\citep{leroy89}. Magnetic configurations with either a dipped field or a helically coiled field
have been invoked to explain the equilibrium and stability of filaments, which leads to three basic
filament models as reviewed by \citet{gilbert01}: the normal polarity dip model
\citep[e.g.,][]{ks57}, the normal polarity flux rope model \citep[e.g.,][]{hirayama85, leroy89},
and the inverse polarity flux rope model \citep[e.g.,][]{kr74, pneuman83, anzer89, lh95}. In
addition, by transporting the core flux into regions of increasingly weak field via shear flows,
\citet{adk94} were able to produce a dipped, inverse-polarity configuration of sheared field lines.
By applying greater shear, \citet{da00} find that magnetic reconnection produces helical field
lines threading the filament. In all these models, the magnetic tension force pointing upward
provides mechanical support for the filament material against gravity. Alternatively,
\citet{karpen01} found that cool plasma can be supported in a dynamic state on flat-topped arcade
field lines and argued that magnetic dips are not necessary for the formation and suspension of
filaments.

The correspondence between the filament chirality and the helicity sign has been controversial due
to different opinions on how the filament is magnetically structured. \citet{rust94} conjectured
that sinistral (dextral) filaments are threaded by right-handed (left-handed) helical fields,
considering that barbs, which are lateral extensions veering away from the filament spine, should
rest at the bottom of the helix. \citet{me94}, on the other hand, noticed that the ends of barbs
are fixed at patches of parasitic polarities (also termed minority polarities), which are opposite
in polarity to the network elements of majority polarity, and suggested that dextral filaments are
right-helical. \citet{cmp05}, however, reported that the barbs terminate over the minority-polarity
inversion line. This lends support to the suggestion that the barb material is suspended
in field line dips which form due to the existence of parasitic polarities. The
latter scenario is consistent with force-free
field models \citep{ad98, aulanier98}, in which the filament spine, the barbs, and the
surrounding fibrils are all modeled as the dipped portions of the field lines. With projection effects,
a continuous pattern of dipped field lines could give the illusion that barbs are made of vertical
fields joining the spine to the photosphere. Projection effects also contribute to the confusion
about the chirality-helicity correspondence of the sigmoidal structures in the corona: the
projection of a single twisted field line includes both forward and inverse S-shapes
\citep{gibson06}. Furthermore, a left-handed flux rope can take either forward or reverse S-shapes
depending on whether it kinks upward or downward \citep{ktt04, Torok&al2010}.

Considerable attention has been given to the eruption of filaments, with the dense filament
material tracing the otherwise invisible progenitor of the coronal mass ejection (CME). Relevant
models can be roughly classified into two categories: those that rely on magnetic reconnections to
remove the tethering field so that the filament can escape \citep{moore01, adk99}; and those in
which the eruption occurs when a flux rope loses equilibrium, due to a catastrophe
\citep[e.g.,][]{vanTend&Kuperus1978, fp95} or due to ideal MHD instabilities
\citep[e.g.,][]{hp79, kt06}. While the torus instability almost certainly plays a role
\citep{Liu2008, aulanier10, fan10}, the occurrence of the helical kink mode has recently been
quite controversial. It is motivated by observations of writhed
eruptive structures \citep[e.g.,][]{rk94, rk96, ji03, rcz03, rl05, alg06, lag07, la09, cho09,kk10},
most of which are filaments. It is also supported by successful MHD
numerical modeling of key properties of eruptions, e.g., writhe, rise profile, and sigmoidal
features, and of specific eruptive events \citep[e.g.,][]{fg04, fan05, fan10, gf06, tkt04, tk05,
kliem10}. On the other hand, signatures of the required amount of twist prior to the eruption
remain difficult to detect \cite[e.g.,][]{Chae2000, Su&al2011}. Eruptions
driven by the kink instability also provide an opportunity to determine the helicity sign of the
filament. Due to helicity conservation, an unstable flux rope only writhes into a kink of the
same handedness, which can be inferred from the writhing motion of the filament axis
\citep[e.g.,][]{green07}, especially if stereoscopic observations are available.

In this paper we address these issues in analyzing the observations of an eruptive
filament which showed strong writhing motions preceded by unwrithing motions. The filament had a
special ``double-decker'' configuration and formed a hard X-ray (HXR) sigmoidal source of coronal
origin between the rapidly rising upper branch and the stable lower branch, which provides some
unique insight into the physics of solar filaments and their eruption. The observations and data
analysis are presented in Section~\ref{s:observations}. Interpretations are discussed in
Section~\ref{s:interpretation}. Section~\ref{s:summary} summarizes the main results.

In Paper~II \citep{Kliem&al2012} we consider the existence, stability, and instability of equilibria containing two
vertically arranged force-free flux ropes in bipolar external field, which further corroborates our
interpretations based on the present data analysis and suggests new models for partial eruptions.

\section{Observations and Data Analysis} \label{s:observations}

\subsection{Instruments and Data Sets}
\begin{figure}\epsscale{1}
\plotone{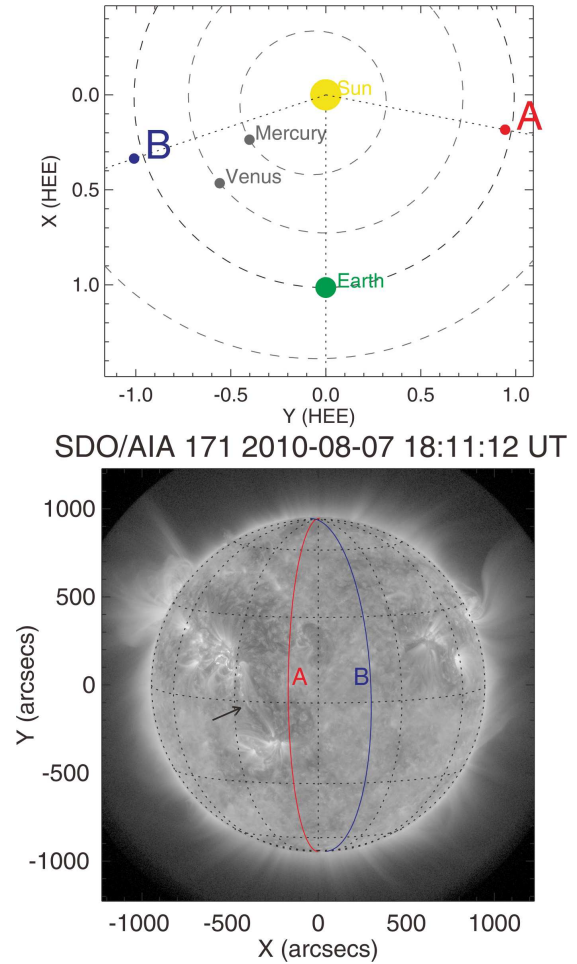} \caption{{\it Top}: positions of the pair of \sat{stereo} satellites in
the Heliocentric Earth Ecliptic (HEE) coordinate system at about 18:00 UT on 2010 August 7; {\it
Bottom}: red and blue curves indicate the limb positions of the \sat{stereo} ``Ahead'' and
``Behind'' satellites, respectively, on an \sat{sdo} image taken near the onset of the eruption.
The arrow marks a group of transequatorial loops connecting
AR~11093 in the northern hemisphere, where a sigmoid is visible, and AR~11095 in the south.
\label{pos}}
\end{figure}

The key data sources in this study include the EUV imaging instruments onboard the
\textsl{Solar Dynamic Observatory} (\sat{sdo}) and the \textsl{Solar Terrestrial Relations
Observatory} (\sat{stereo}; \citealt{kaiser08}), and the \textsl{Reuven Ramaty High Energy
Solar Spectroscopic Imager} (\sat{RHESSI}; \citealt{lin02}).

The Atmospheric Imaging Assembly \citep[AIA;][]{lemen11} onboard \sat{sdo} takes EUV/UV images at
multiple wavelengths with a resolution of $\sim$1$''.2$ and a cadence of 12 seconds for each
individual wavelength, covering an unprecedentedly wide and nearly continuous temperature range. AIA
Level-1 data are further processed using the standard SSW procedure \texttt{AIA\_PREP} to perform
image registration.

The Extreme-UltraViolet Imager (EUVI; \citealt{wuelser04}) of the Sun Earth Connection Coronal and
Heliospheric Investigation (SECCHI; \citealt{howard08}) imaging package onboard both \sat{stereo}
satellites provides three passbands, namely, 171~\AA\ (\ion{Fe}{9}), 195~\AA\ (\ion{Fe}{12}) and
304~\AA\ (\ion{He}{2}). The top panel of \fig{pos} shows the positions of the \sat{stereo}
``Ahead'' and ``Behind'' satellites (hereafter referred to as \sat{stereo-a} and \sat{stereo-b},
respectively) in the X-Y plane of the Heliocentric Earth Ecliptic (HEE) coordinate system. In the
bottom panel, the red and blue arcs mark the corresponding limb positions of the \sat{stereo}
satellites in the AIA image obtained at 18:11~UT on 2010 August~7. The active region of interest,
NOAA Active Region 11093, was located at N12E31, connecting with AR 11095 (S19E20) in the southern
hemisphere through a group of transequatorial loops as indicated by an arrow.

\subsection{Pre-eruption Configuration} \label{ss:obs_configuration}
\begin{figure}\epsscale{1}
\plotone{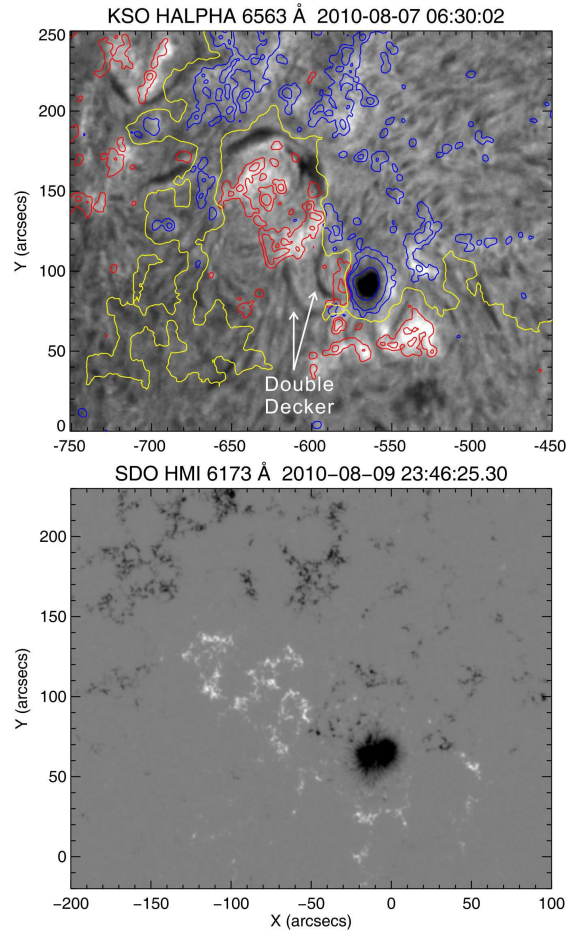} \caption{H$\alpha$ image taken by Kanzelh\"{o}he Solar Observatory (KSO;
\textit{top} panel), overlaid by an HMI line-of-sight magnetogram taken at the same time. Contour
levels indicate the magnitude of magnetic flux density at 50, 200, and 800 G for positive (red)
and negative (blue) polarities. Yellow contours indicate the major PIL of the active region.
The \textit{bottom} panel shows a line-of-sight magnetogram taken two days later when the
sunspot was close to the disk center. \label{ha}}
\end{figure}

The filament located in AR 11093 was composed of two branches, hereafter referred to as the
\emph{lower branch} and the \emph{upper branch}. As shown in \fig{ha}, the lower branch was
aligned along the PIL as filaments usually do, while the upper branch was projected onto the
region of positive polarities, owing to the fact that it was located high in the corona
(Section~\ref{ss:3d}). The southern ends of both branches were apparently rooted in the penumbra
of the sunspot that was of negative polarity. (Note that due to the projection effect the field
in the eastern periphery of the penumbra possessed a positive component along the line of sight. This is clearly demonstrated in the line-of-sight magnetogram taken two days later (bottom panel of \fig{ha}), in which the sunspot was close to the disk center and the positive patch to the east of the sunspot has disappeared.)
Hence, the field direction along the filament axis must be pointing southward, spiraling
counterclockwise into the sunspot. Therefore, both branches of the filament were dextral,
according to the definition of the filament chirality \citep{martin98}. This is consistent with
the empirical hemispheric chirality rule of filaments.

\subsection{3D Reconstruction and Analysis} \label{ss:3d}
\begin{figure*}\epsscale{1}
\plotone{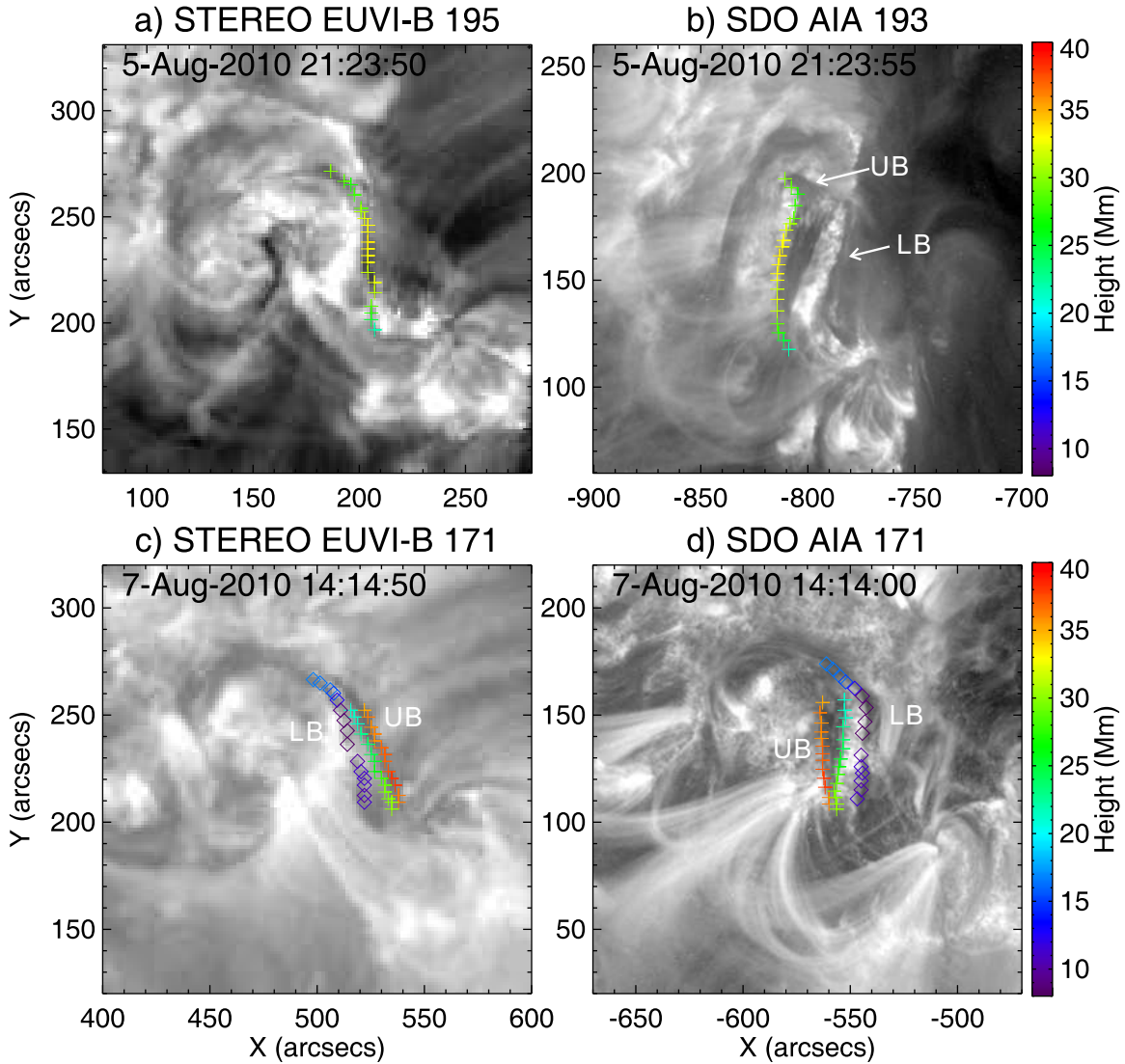} \caption{Three-dimensional reconstruction of the filament height using the
tie-point method. Points on the upper branch (UB) are indicated by crosses and those on the
lower branch (LB) by diamonds. The heights of these points above the solar surface in Mm are
color-coded as indicated by the color bar on the right. \label{stereo}}
\end{figure*}

By using a pair of EUV images taken from different perspectives, the three-dimensional
location of the filament under investigation can be derived by a triangulation technique called
\emph{tie point} \citep{inhester06}. This is implemented in an SSW routine, \texttt{SCC\_MEASURE},
by W.~Thompson and has been utilized to obtain the three-dimensional shape and height of filaments
observed by \sat{stereo} \citep[e.g.,][]{li10, xjw10, seaton11, Bemporad&al2011, thompson12}.
In our case, \sat{stereo-b} images are
to be paired with earth-view images provided by \sat{sdo}, as the filament was occulted from the
\sat{stereo-a} view (\fig{pos}). In EUVI images, the filament is more clearly defined with better
contrast in 171 {\AA} than in 195 {\AA} or 304 {\AA}, but 171 {\AA} images are often not available,
suffering from much lower cadence (2 h) than the other two channels (usually 5--10 min).

The two observers, \sat{sdo} and \sat{stereo-b}, and the point in the solar corona to be triangulated
define a plane called the epipolar plane. For each observer, the epipolar plane intersects
with the image plane in the epipolar line. The basis for triangulation is known as the epipolar
constraint \citep{inhester06}, i.e., a point identified on a certain epipolar line in one image
must be situated on the corresponding epipolar line in the other image. Matching features that appear
on a pair of epipolar lines therefore establishes a correspondence between pixels in each image, and the
reconstruction is achieved by tracking back the lines of sight for each pixel, whose intersection
in the same epipolar plane defines a unique location in three-dimensional space.

\fig{stereo} shows two pairs of EUV images from AIA onboard \sat{sdo} and from EUVI onboard
\sat{stereo-b} taken on 2010 August 5 and 7. On August 5, both branches of the filament can be
clearly seen in the AIA image (\fig{stereo}(b)), while only a single thread-like structure of
reverse S-shape was displayed by the corresponding \sat{stereo-b} image (\fig{stereo}(a)). For \sat{stereo-b}
the filament was located close to the disk center, suggesting that the upper branch was located right
above the lower branch. Hence, we will refer to the configuration as a ``double-decker
filament.'' By triangulation, we obtain the height of the upper branch, which was at
about $30\pm3$ Mm above the solar surface. The reconstruction error is $\sim w/\sin\alpha\simeq2.6$
Mm, where $w\simeq2.5$ Mm is the width of the observed filament thread in the \sat{stereo-b} image and
$\alpha\simeq71^\circ$ is the separation angle between \sat{stereo-b} and \sat{sdo} on August 5.

On August 7, the two branches were visible in both the \sat{stereo-b} and \sat{sdo} views
(\fig{stereo}(c) and (d)). About 4 hours later, the upper branch in the \sat{sdo} view moved
eastward and then erupted, while in the \sat{stereo-b} images, one can see
that the branch in the west moved westward and then erupted. Hence we are able to match the two
branches unambiguously in the image pair, \fig{stereo}(c) and (d). For the upper branch, the
triangulation results show that its upper rim was located at about $36\pm2$ Mm above the surface, 6
Mm higher than it was on August 5, corresponding to an average rising speed of about \speed{0.1}.
Its lower rim was $25\pm3$ Mm high, hence the vertical extent of the upper branch was about
$11\pm4$ Mm. For the lower branch, its upper rim had a height of about $12\pm3$ Mm, which was
separated from the lower rim of the upper branch by a distance similar to the vertical extension of
the upper branch.

\subsection{Filament Rise and Eruption} \label{ss:rise&eruption}
\begin{figure*}\epsscale{1}
\plotone{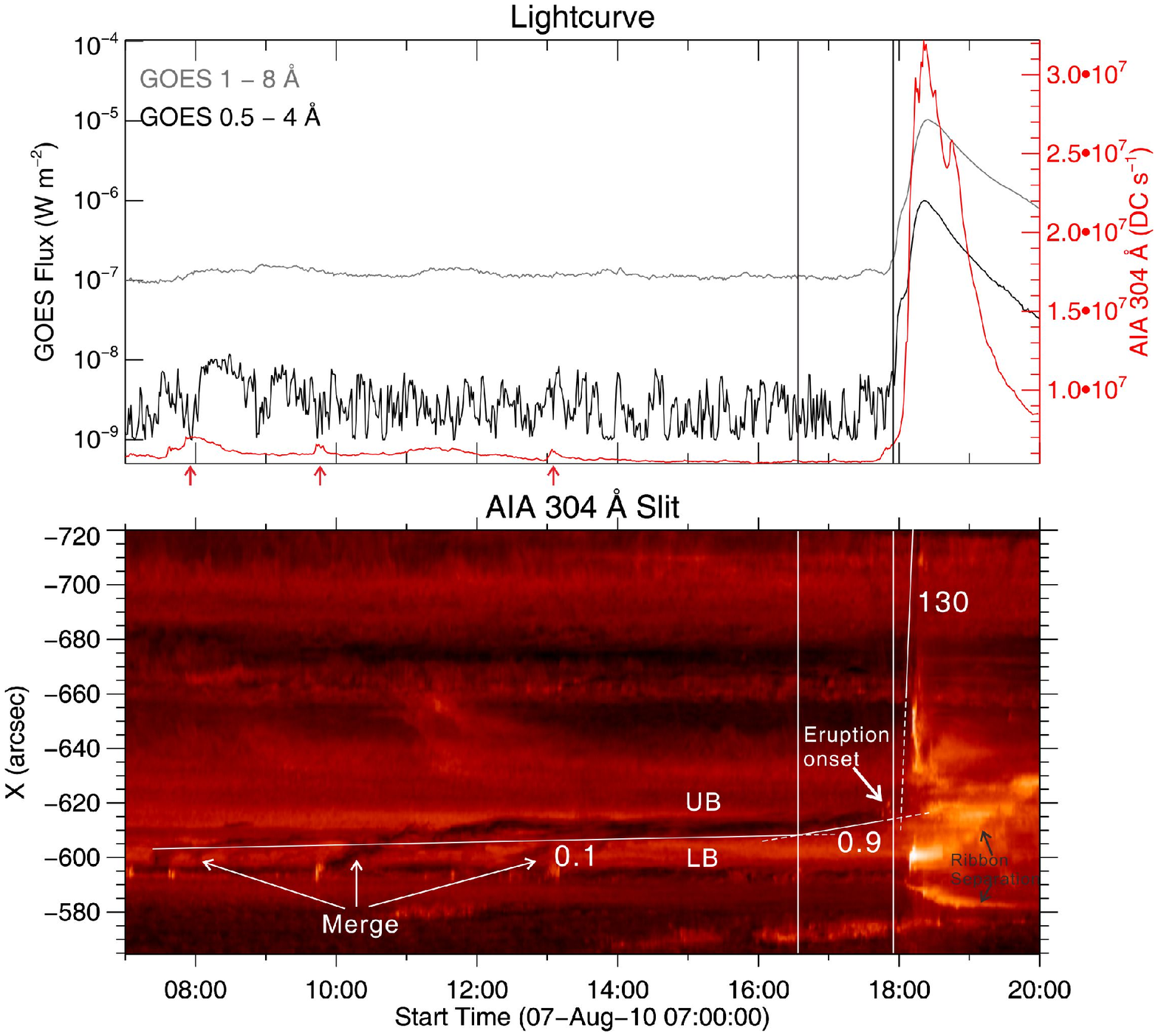} \caption{Intermittent merging process observed in AIA 304 \AA.
\textit{Top panel}: \sat{goes} and AIA lightcurves. The AIA lightcurve (red) is obtained by
integrating over the green box in \fig{erupt}(a). Three bumps in the lightcurve corresponding to
the heating at the start of each merging process are marked by arrows in the bottom.
\textit{Bottom panel}: a space-time diagram (stack plot) obtained from the slit displayed in the
accompanying animation of AIA 304~{\AA} images for the same time interval, with all images
being registered with the first image at 07:00:10 UT. The intensities in the space-time diagram
are displayed in logarithmic scale. The first two sloping lines are adjusted to the western
edge of the upper branch, which is better defined than its eastern edge, and the third sloping
line follows the eruptive feature, which briefly swept across the slit from 18:07 to 18:17 UT.
Dashed lines are simply extensions of the solid lines which outlines the corresponding features.
The implied velocities are quoted in km\,s$^{-1}$. The two vertical lines mark the crossing of
the first two sloping lines and the onset of the \sat{goes} flare, respectively. \label{slit}}
\end{figure*}

\begin{figure*}\epsscale{1}
\plotone{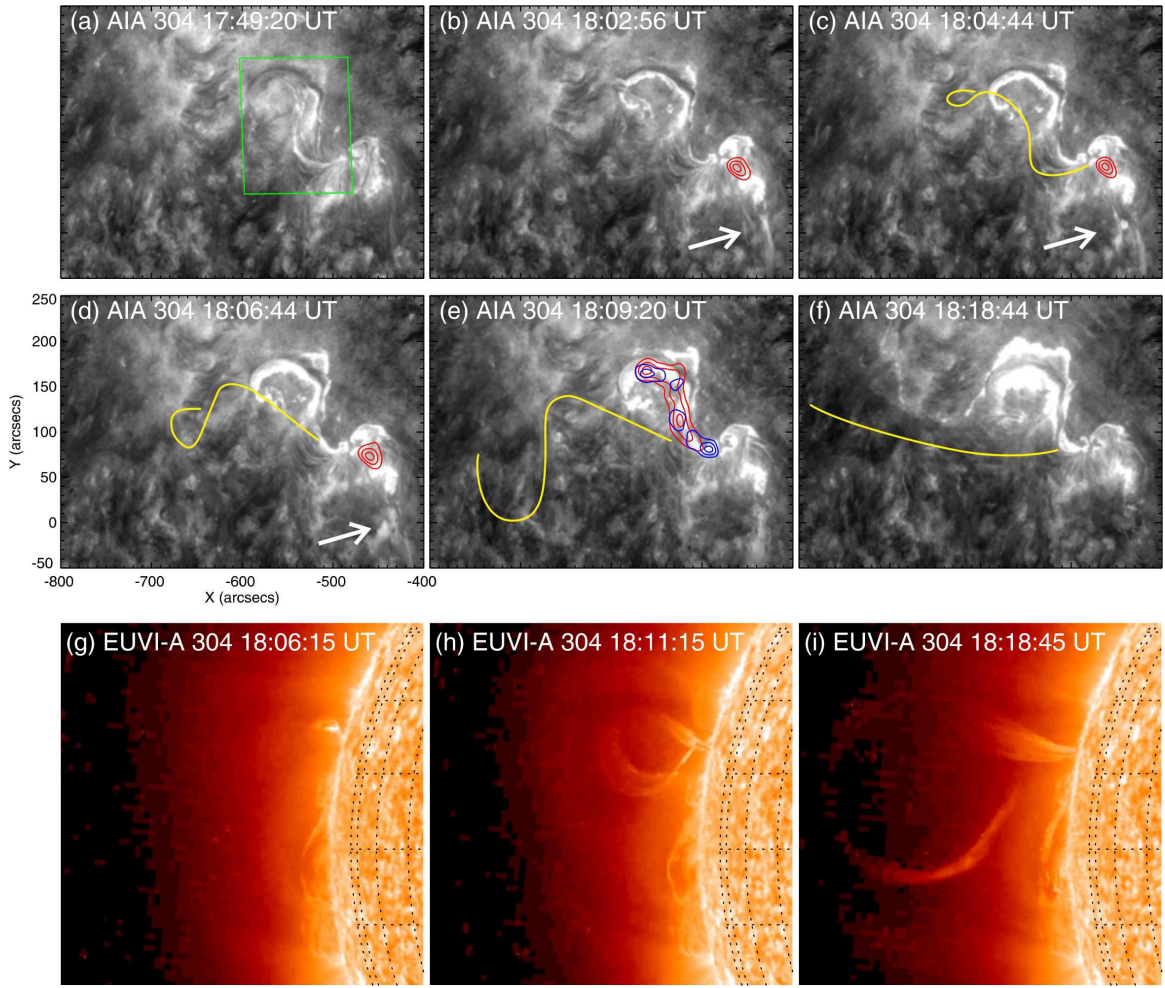} \caption{Filament eruption observed in \ion{He}{2} 304 {\AA} in two
different viewpoints. Panels (a)--(f): snapshots of AIA 304 {\AA} images, overlaid by contours
indicating \sat{rhessi} emission at 3--9 keV (red) and 15--40 keV (blue), respectively.
Contour levels are set at 50\%, 70\% and 90\% of the maximum brightness of each individual image.
Yellow curves highlight the upper branch of the filament which was too diffuse to be seen in
static images but visible in the animation accompanying the figure; in the animation, the left
panel shows an original AIA 304 \AA\ image and the right panel shows the same image enhanced with
a wavelet technique). White arrows mark the flare spay propagating along the transequatorial
loops connecting to AR 11095 in the southern hemisphere. Panels (g)--(i): 304 {\AA} images
obtained by the Ahead satellite of \sat{stereo}. The green box in Panel (a) marks the
integration area to obtain the AIA 304~{\AA} lightcurve in Figure~\ref{slit}. It is slightly
slanted due to differential rotation, as the box is defined in the image at 07:00:08 UT.
\label{erupt}}
\end{figure*}

The evolution of the active region's photospheric field in the course of the 1--2 days during
which it was well visible before the eruption was mainly characterized by the moat flow around
the sunspot and ongoing gradual dispersion of the positive polarity flux in the northern hook of
the filament. Episodes of significant new flux emergence, shearing, or cancellation were not
observed. Weak flux cancellation proceeded at the north-south directed section of the PIL under
the middle of the filament. The two filament branches did not show strong changes in position
either. However, there were quite dramatic signs of interaction between them, which appeared to
be the most significant changes associated with the filament system, in addition to the weak flux
cancellation under its middle part. Episodes repeatedly occurred in which a dark thread within
the lower branch was heated, brightened, lifted upward, and merged with the upper branch,
apparently cooling down again. To show this evolution, we place a virtual ``slit'' across the
center of the double-decker filament in the AIA 304~{\AA} images and generate a stack plot from
these data. The slit is taken to be 260 by 10 pixels spanning from $-719.7''$ to $-564.3''$ in
the east-west direction and from $119.1''$ to $124.5''$ in the north-south direction (see the
animation accompanying \fig{slit}). The intensities are summed in vertical direction, and the
solar rotation and differential rotation are removed. The resulting space-time diagram is plotted
in \fig{slit} in the range 07:00:10 to 19:59:10 UT. Three exemplary mergers occurring within 10
hours prior to the eruption are visible. Each merger corresponds to a small bump in the light
curve of the AIA 304~\AA\ channel (top panel of \fig{slit}), which is obtained by integrating
over a region enclosing the filament (\fig{erupt}(a), green box). The bumps manifest the heating
of the filament, due to the reconnection associated with flux transfer from the lower to the
upper branch. Assuming dominantly horizontal field direction in the filament, the mass transfer
implies a corresponding flux transfer.

The merging episodes cannot be associated with significant changes in the position of the
branches, but they occurred in a long-lasting phase of very gradual displacement (likely gradual
rise) of the upper branch, as indicated by the first of the three sloping lines in \fig{slit}.
The slope of the line represents a velocity of 0.1~km\,s$^{-1}$ in the plane of the sky,
consistent with the average rise velocity since August 5 ($\sim\!0.1$~km\,s$^{-1}$) estimated
from the three-dimensional reconstruction in Section~\ref{ss:3d}. Both the transfer of flux into
the upper branch and the weakening of the overlying field by the gradual cancellation are
plausible causes of the indicated very gradual rise of the branch.

Following the mergers, the upper branch appeared thicker and darker, presumably due to the mass
loading effect. A few other factors could also affect the observed absorption, such as Doppler
shift, change in perspective, shift of the filament structure, and condensation of coronal
plasma. It is difficult to evaluate how these effects impact on filament darkness and
thickness in relation to the mass loading, which is the most obvious factor.

About 1.5~hours prior to the eruption, the displacement of the upper branch accelerated slightly
to reach more typical velocities of a slow-rise phase; the slope of the second line in \fig{slit}
corresponds to about 0.9~km\,s$^{-1}$. Whether this change occurred abruptly at the crossing time 
(16:34 UT) of the lines or gradually, and whether it was due to external influences or changing
conditions in the filament, cannot be determined from the available data. No significant external
influence on the filament can be discerned near this time.

Threads in the upper branch lightened up in the EUV from about 17:42~UT. The resulting mix of
dark and bright structures strongly increased the visibility of the internal motions in both
branches (see the animation accompanying \fig{erupt}). At earlier times, motions directed toward
the sunspot could be seen throughout the lower branch and in the southwest half of the upper
branch. With the onset of the brightening, the motions in the upper branch intensified, becoming
faster in the southwest half. Additionally, motions toward the other end of the branch became
very visible in the northeast half (this may be a true enhancement or be due to the improved
perception). The motions toward the ends of the upper branch continued into the onset phase of
the eruption.

The eruption of the upper filament branch commenced in relatively close temporal association with
the soft X-ray (SXR) flare. It can be seen from \fig{slit} that the brightening and acceleration of the
upper branch (marked by an arrow) started slightly earlier (about 10 min) than the onset of
the \sat{goes}-class M1.0 flare at about 17:55~UT. Due to the acceleration, the crossing time of
the second and third sloping lines slightly lagged (about 7 min) behind the onset of flare. Within
about two minutes, also a flare spray was ejected from the vicinity of the sunspot, but into a
different direction (\fig{erupt}). The flare spray propagated along the transequatorial loops
connecting with AR~11095 in the southern hemisphere (see also \fig{pos}). The lower branch of the
filament, on the other hand, obviously remained stable at its original location, flanked by a
pair of conjugate flare ribbons in 304~{\AA} separating from each other (\fig{erupt}, see also
\citealt{vemareddy11} for a study of the ribbon separation in relation to the filament eruption).
The eruption evolved into a fast CME detected by the \sat{soho}/LASCO and \sat{stereo}/COR
coronagraphs.

\begin{figure*}\epsscale{1}
\plotone{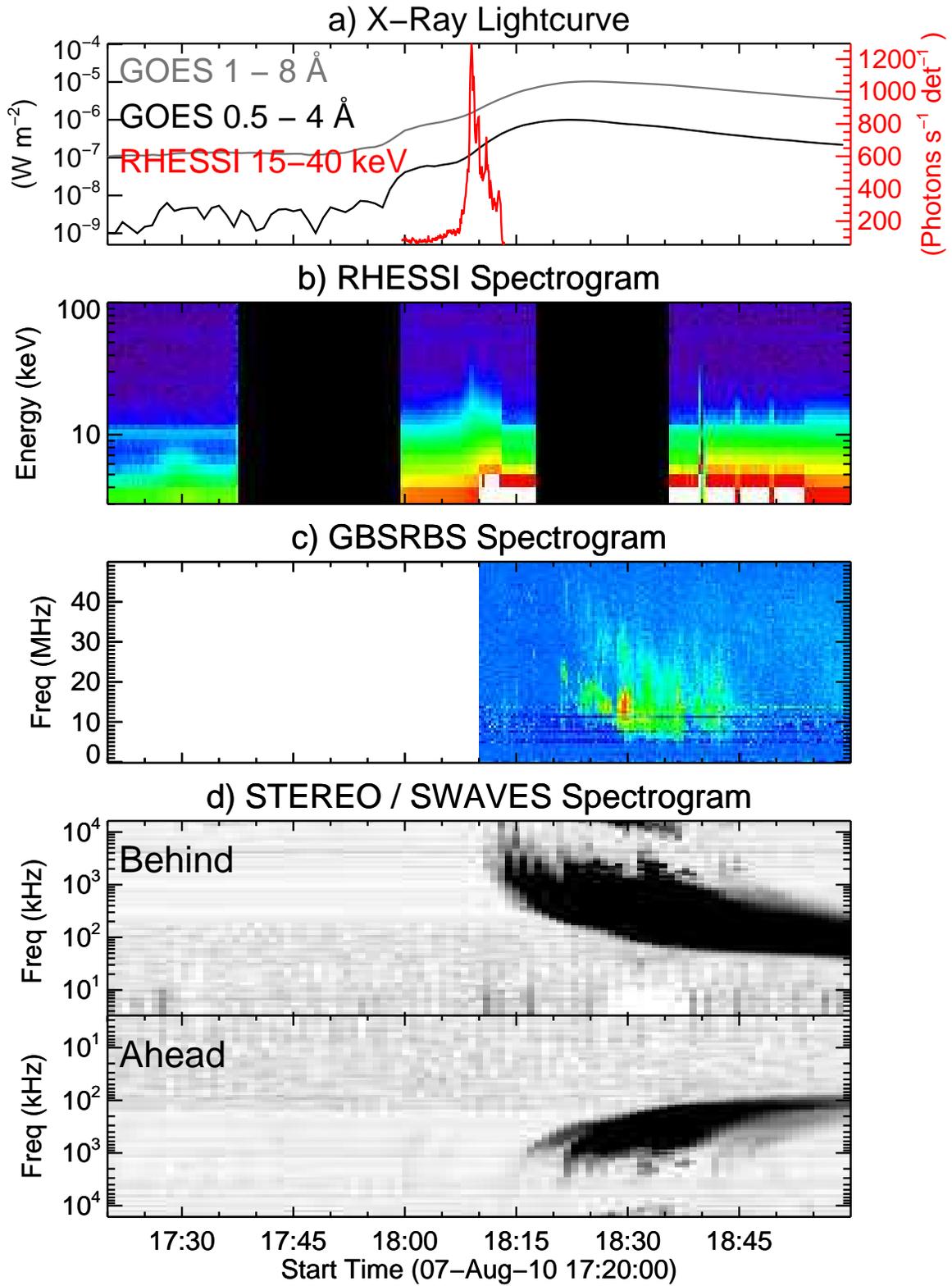} \caption{Time evolution of the filament eruption in X-rays and radio.
\label{lc}}
\end{figure*}

The upper filament branch experienced a dramatic change of shape in the course of its eruption.
Originally, it had a strong reverse S shape following the PIL (Figures~\ref{stereo} and
\ref{erupt}). The AIA images in \fig{erupt} first indicate a straightening of the branch
(Panel~(c)), followed by a strong writhing into a forward S shape (Panels (d)--(e)).
With yellow curves, we highlight the shape of the branch, which was
too diffuse to be clearly seen in static images, but is visible in the animation of these images
accompanying \fig{erupt}. Both the reverse S shape of the low-lying filament branches and the
forward S shape of the erupted, high arching upper branch are signatures of left-handed writhe
\citep{Torok&al2010}, indicating left-handed chirality for this dextral filament.

These considerations are supported by the \sat{stereo} data.
In the \sat{stereo-a} 304 {\AA} images in which the upper branch of the filament was above the east limb
from 18:06 UT, a leg-crossing loop configuration developed (\fig{erupt}(g)--(i)), indicative
of strong writhing.
By examining the animation of AIA 304 {\AA} images, one can see that the filament leg that was fixed
at the sunspot penumbra was more perceptible than the other leg which became too tenuous to be seen
by 18:05~UT. Assuming a similar visibility for the corresponding \sat{stereo-a} images in the
same wavelength, we identify the filament leg that was apparently attached to the surface in the
\sat{stereo-a} view (\fig{erupt}(h)--(i)) with the leg fixed at the sunspot penumbra in the earth view
(\fig{erupt}(a)--(f)). Since this footpoint is closer to the observer, the filament traces a
left-handed helical curve, consistent with the left-handed writhe inferred from the
counterclockwise rotation of the filament's upper section in the AIA images (\fig{erupt}(c)--(e)).

\subsection{Associated Flare}\label{ss:flare}
\begin{figure*}\epsscale{1}
\plotone{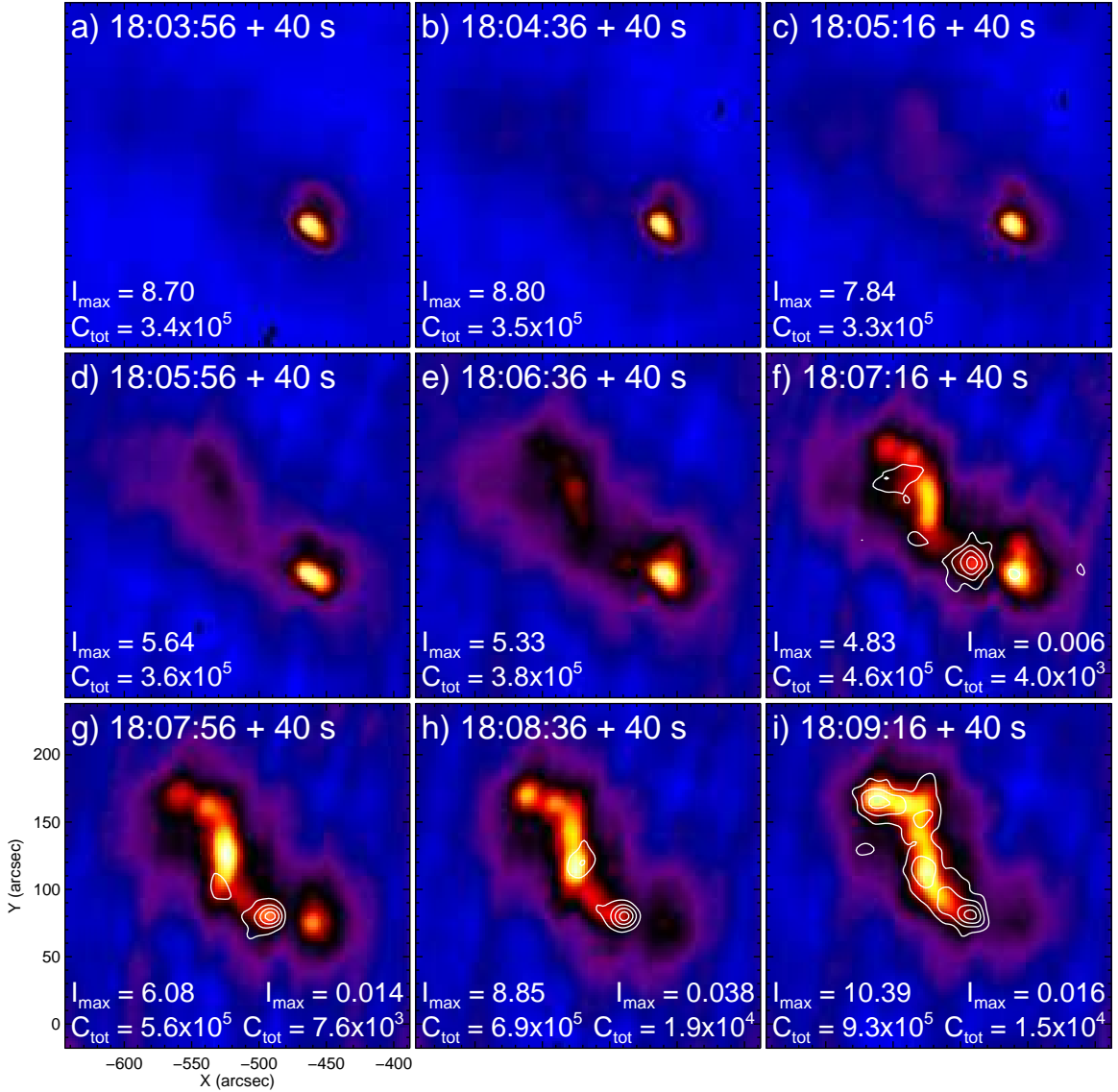} \caption{Evolution of flare morphology in HXRs. The figure shows a series of
\sat{rhessi} CLEAN images at 3--9 keV, overlaid by contours at 50\%, 70\% and 90\% of the maximum
brightness, $I_{max}$ (photons cm${}^{-2}$ s${}^{-1}$ arcsec${}^{-2}$) of each image obtained for
the same time interval at 15--40 keV. $I_{max}$ is given in the bottom left (right) of each panel
for the 3--9 (15--40) keV energy ranges; also quoted is $C_\mathrm{tot}$, the total counts accumulated by
Detectors 3--8. \label{hsi}}
\end{figure*}

\begin{figure*}\epsscale{1}
\plotone{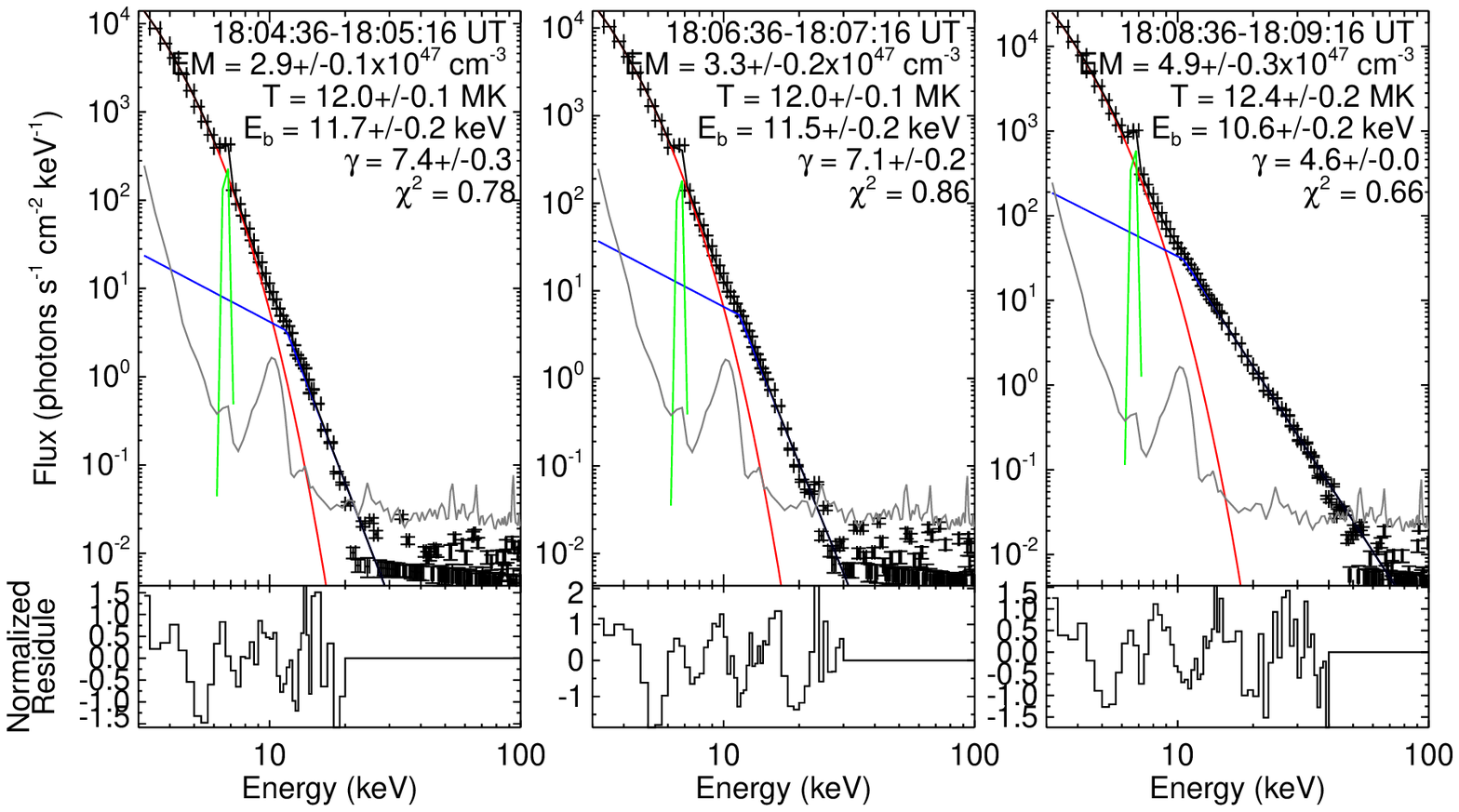} \caption{\sat{rhessi} spatially integrated, background-subtracted spectra
and the corresponding spectral fits. The three time intervals correspond to the three panels in
the middle column of \fig{hsi}. The red line indicates an optically thin thermal bremsstrahlung
radiation function, the blue line a broken power-law function, the green line a Gaussian
function, and the grey line the background. The fitting parameters are shown in each panel,
including emission measure ($EM$), temperature ($T$), break energy of the broken power-law
function ($E_b$), spectral index ($\gamma$) above the break energy, and Chi-square ($\chi^2$) of
each fitting, assuming a system uncertainty of 5\%. The spectral index below the break energy is
fixed at 1.5. The maximum energy to fit is chosen automatically by the software based on the
spectral values with respect to the background. Fitting residuals normalized to the 1$\sigma$
uncertainty of the measured flux are shown at the bottom. Detectors 1, 4, and 6 are used.
\label{spectra}}
\end{figure*}

X-ray flaring activity associated with the eruption commenced as a microflare at 17:30~UT,
detected by \sat{rhessi} in the energy range $\approx$\,3--9 keV (\fig{lc}(b)). In the
\sat{goes} soft X-ray flux, the
signal was very weak, barely above the background noise (\fig{lc}(a)). \sat{rhessi} images at
3--9~keV show a compact source located to the west of the sunspot. The source location and morphology
was almost identical to the early phase of the \sat{goes}-class M1.0 flare starting at about
17:56~UT (\fig{erupt}(b)--(d)), suggesting that the magnetic configuration stayed
largely unchanged in this interval. No specific change in the filament can be associated with
the microflare. It appears that this tiny event did not play any significant role in the evolution
toward the eruption.

The spatial and spectral evolution of the M1 flare is shown in Figures~\ref{hsi} and
\ref{spectra}, respectively. Throughout the flare interval, the spatially integrated spectra
can be well fitted by an isothermal, exponential component below about 10~keV (red line) and a
nonthermal, power-law component above about 15~keV (blue line) plus a narrow Gaussian
component emulating the iron-line complex at 6.7~keV (green line). Toward the flare peak, the
power-law section hardened significantly, and the flare morphology underwent a drastic change
as the upper branch of the filament rose and writhed. Until 18:08~UT, the compact
source to the west of the sunspot was still visible, but new emission began to develop to the east
of the sunspot from 18:06~UT, whose shape roughly followed the curved filament. At the peak of the
nonthermal emission at about 18:09~UT (\fig{lc}(a)), a sigmoidal HXR source was fully developed
(\fig{hsi}(h) and (i)), outshining the compact source to the west of the sunspot.

In the dynamic spectrograms obtained by the Green Bank Solar Radio Burst Spectrometer
(GBSRBS\footnote{\url{http://gbsrbs.nrao.edu/}}) on the ground (\fig{lc}(c)) and the WAVES
instrument onboard \sat{stereo} \citep{bougeret08} (\fig{lc}(d)), one can see a fast-drift burst
(type~III) starting at about 08:12~UT near 35~MHz, indicating the escape of some nonthermal
electrons along open field lines, presumably due to the interaction of the filament field with the
surrounding field. One can also see a slow-drift (type~II) burst commencing at about 18:20~UT
near 50~MHz, which reveals the formation of a large-scale coronal shock by the fast CME.

\subsection{Coronal HXR Sigmoid}\label{ss:sigmoid}
\begin{figure*}\epsscale{1}
\plotone{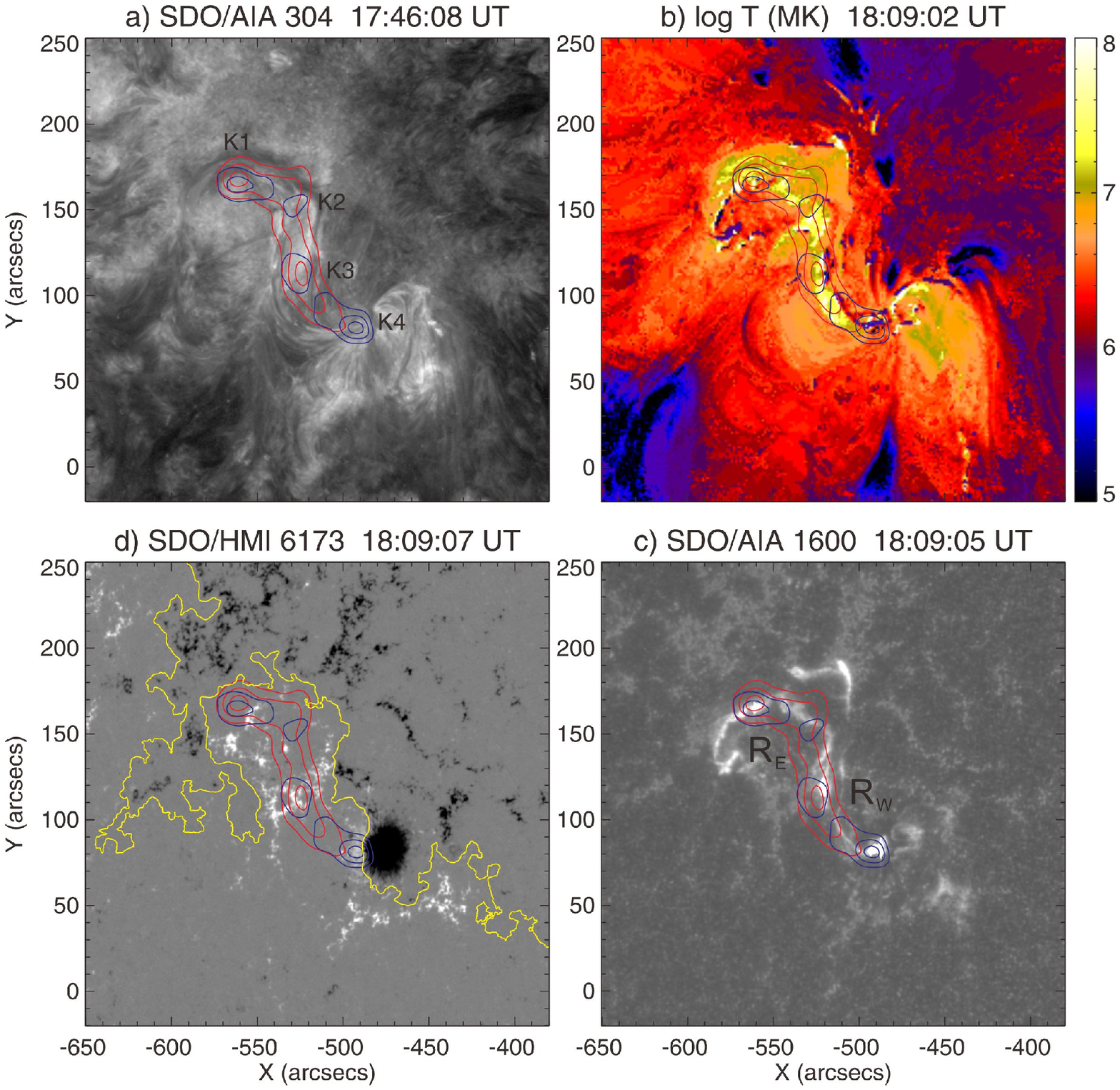} \caption{Sigmoidal HXR source in relation to (a) the double-decker
filament, (b) flaring loops in the corona,  (c) flare ribbons in the chromosphere, and (d)
photospheric line-of-sight magnetic field. Panel (b) is a temperature map obtained with the DEM
method using six AIA coronal channels. Red and blue contours show the 3--9 and
15--40~keV emission, respectively, integrated over 18:09:16--18:09:56~UT.
\label{sigmoid}}
\end{figure*}

The HXR nonthermal emission at 15--40 keV was composed of four kernels, $K1$--$K4$
(\fig{sigmoid}(a)), at the flare peak. Overall it took a reverse S-shape, and $K1$--$K3$ were
co-spatial with the thermal source (3--9 keV) that possessed a continuous, reverse S-shape
passing over the PIL. This thermal sigmoidal source was roughly projected to the gap between the
two filament branches in the AIA 304 \AA\ image (\fig{sigmoid}(a)), and was co-spatial with a hot
ridge in the temperature map (\fig{sigmoid}(b)) which is obtained with the differential emission
measure (DEM) method utilizing the six AIA coronal wavelengths \citep[131, 171, 193, 211, 335,
and 94 \AA;][]{ab11}. The temperature of this hot ridge is consistent with the isothermal
temperature of the SXR emission derived from spectral fitting ($\sim$12 MK;
\fig{spectra}), confirming that the flaring plasma was observed by both \sat{rhessi} and AIA.
Thermal sources are usually coronal sources, not footpoint sources. The continuous structure,
passing over the PIL, clearly favors a coronal source in the present event as well. The hot ridge
must show plasma in the flare current sheet above the arcade of flare loops or the tops of the
flare loops which are known to often have the highest brightness in the arcade.

\begin{figure*}\epsscale{1}
\plotone{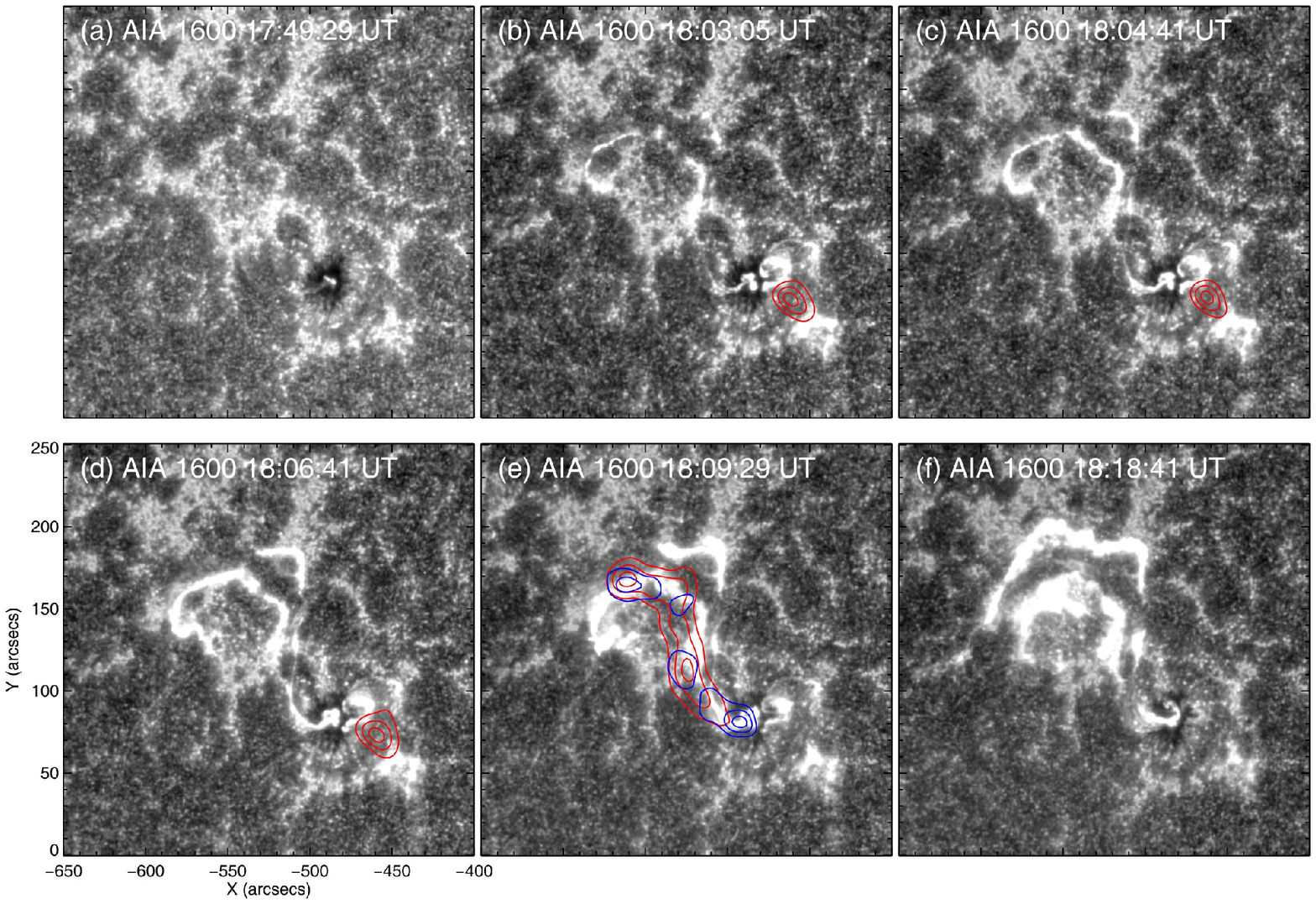} \caption{UV flare ribbons in relation to HXR emission represented by
contours (same as in \fig{erupt}). \label{ribbon}}
\end{figure*}

\begin{figure*}\epsscale{1}
\plotone{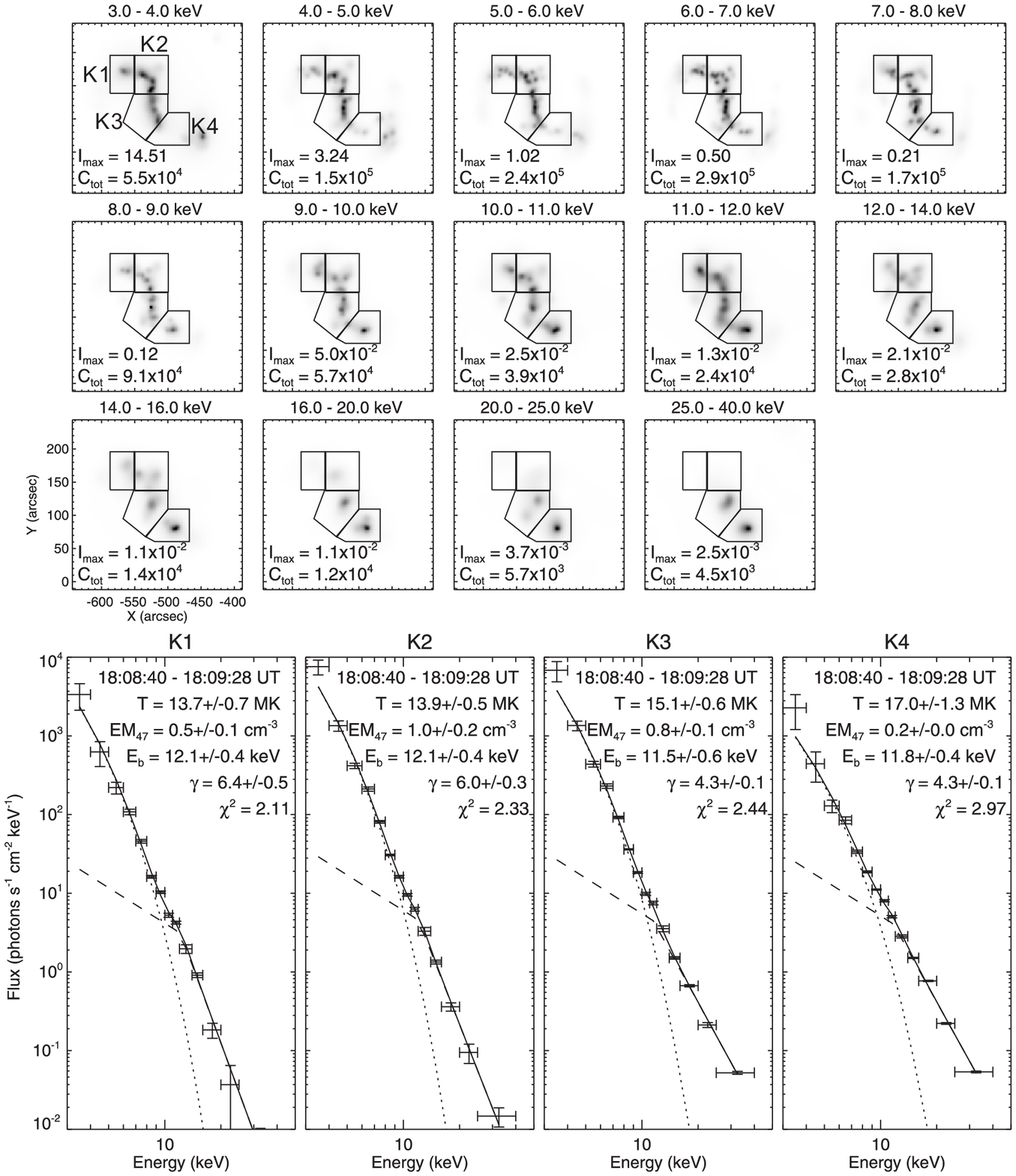} \caption{Imaging spectroscopy for a 48 s interval at the peak of the HXR
burst at about 18:09 UT on 2010 August 7. Pixon images are made with detectors 2--8 in 14 energy
bins from 3--40 keV. Indicated in the bottom left corner of each image are the maximum
brightness, $I_{max}$ (photons cm${}^{-2}$ s${}^{-1}$ arcsec${}^{-2}$), of each individual Pixon
map, and the total counts accumulated by the detectors used, $C_\mathrm{tot}$. The spatially
resolved spectrum for each region as marked by polygons is fitted with an isothermal function
(dotted line) plus a broken power-law function (dashed line) with the spectral index below the
break energy being fixed at 1.5. Resultant fitting parameters, as in Figure~\ref{spectra}, are
given for each spectrum. Emission measures given as EM${}_{47}$ are to be multiplied by
${10}^{47}$. \label{imspex}}
\end{figure*}

From the above observations, we conclude that the sigmoid featured by SXR emission was of
coronal origin. But were the nonthermal kernels, $K1$--$K4$, coronal or footpoint sources?
First of all, we believe that these kernels are unlikely ``ghost'' images due to the pulse
pileup effect \citep{smith02, hurford02}, as the fractional livetime is relatively high (87.96\%
at the peak of the nothermal emission) and the countrate is relatively low (the peak countrate at
12--25 keV is $\sim200\ \mathrm{s}^{-1}\ \mathrm{det}^{-1}$ with no attenuators in front of the
detectors). The spatially resolved spectra of $K3$ and $K4$ (\fig{imspex}) were significantly
harder than the other two kernels, making these the strongest candidates for footpoint sources.
The source $K4$ coincided with the southern end of the western, negative-polarity ribbon $R_W$
(Figures~\ref{sigmoid}(c) and \ref{ribbon}), and it was spatially distinct with both the SXR
source and the EUV hot ridge (\fig{sigmoid}(b)), hence it was very likely a footpoint
source. The sources $K1$--$K3$ had projected locations on the positive-polarity side of the PIL.
While the stronger source K1 fell on the eastern, positive-polarity ribbon $R_E$, the centroids
of $K2$ and $K3$ were clearly displaced from the ribbon. (Note that $R_E$ was shorter, ending
near $K3$, while $R_W$ ended at $K4$.) Moreover, both UV ribbons were formed as early as 18:04
UT, before the HXR sigmoid appeared (\fig{ribbon}). This is in contrast to previously reported
cases, in which UV bright kernels tended to be both co-spatial and co-temporal with HXR
footpoints \citep[e.g.,][]{ww01, liuc07, qiu10}, although the latter are often more compact. This
timing, the spatial association of $K2$ and $K3$ with the SXR source and their displacement from
the UV ribbon are all naturally explained if $K2$ and $K3$ were coronal sources. The nature of
kernel $K1$ is less clear: its location is consistent with a footpoint source (although not
inconsistent with a coronal source falling on the ribbon only in projection), while its soft
spectrum, similar to $K2$ (\fig{imspex}), indicates a coronal source. Overall, the HXR sigmoid
observed here is at least partly of coronal origin, different from that reported by \citet{ji08},
which appeared in a typical flare morphology with two conjugate footpoints and a looptop source.

\section{Interpretation} \label{s:interpretation}
      % {Discussion \& Conclusions}

\subsection{Filament Chirality and Magnetic Configuration} \label{ss:configuration}

\begin{figure}\epsscale{1}
\plotone{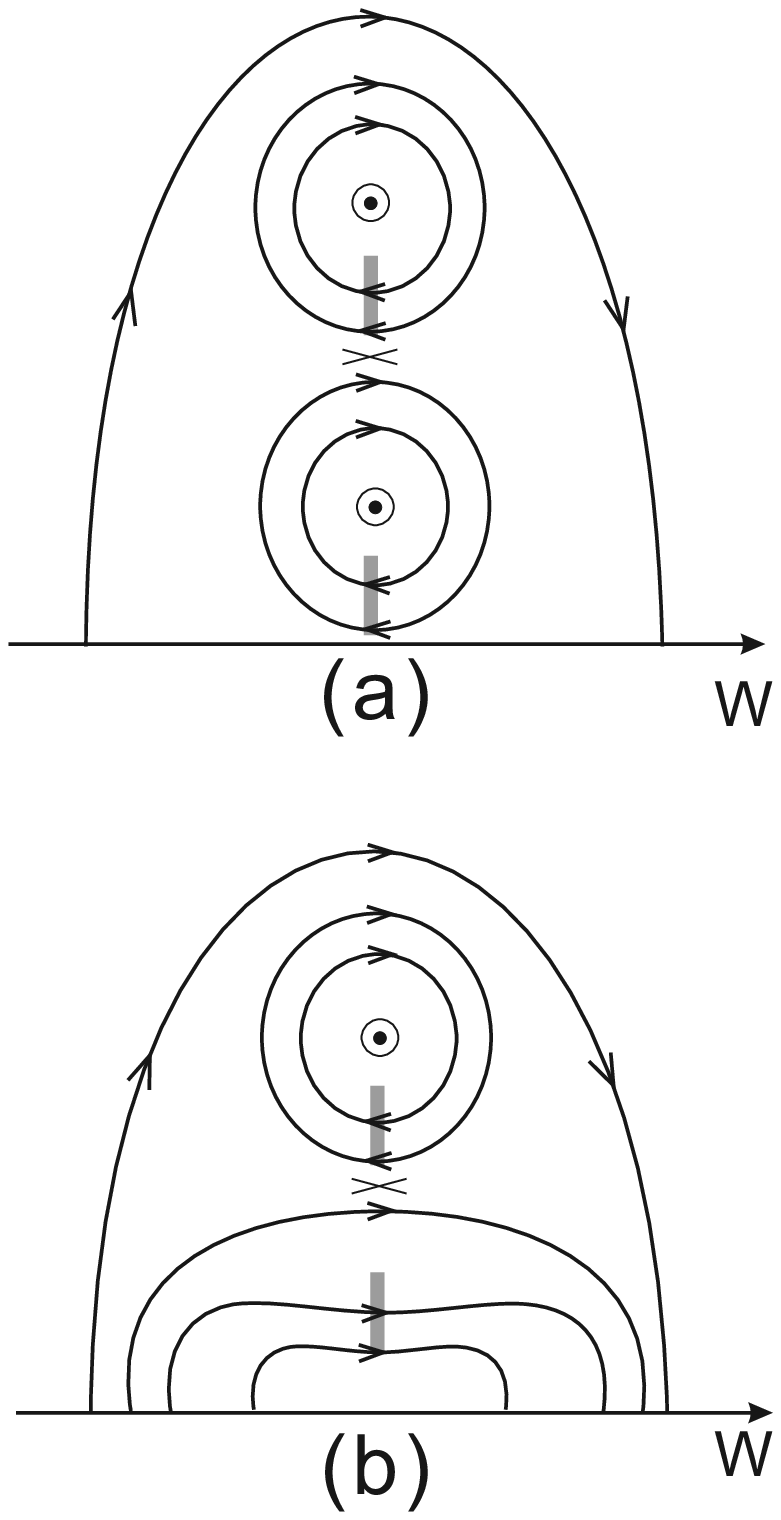} \caption{Cartoon illustrating the cross section of the two suggested
double-decker filament configurations as viewed from the south.
The axial field of both filament branches
points out of the plane. Regions to the east (west) of the filament are associated with
positive (negative) polarity. Slabs of grey color indicate the filament body. \label{cartoon}}
\end{figure}

The writhing motion and the consequent formation of the left-handed kink clearly indicate that the
upper branch of the filament was embedded in a left-handed flux rope. Thus, the inverse polarity
configuration is indicated for this branch. The helicity sign of the lower branch can be
inferred from its magnetic
connections at the ends, which were the same as for the upper branch, and from its interaction with the
upper branch. The MHD simulations by \citet{linton01} showed that two parallel flux tubes with the
same sign of helicity tend to merge while those with the opposite sign of helicity tend to bounce
off each other. Thus, left-handed helicity is indicated for the lower branch as well.

% [, i.e., for the dextral filament as a whole.]

However, the
lower branch could be either embedded in a flux rope with the inverse polarity configuration
(\fig{cartoon}(a)) or in a sheared arcade (\fig{cartoon}(b)) with the normal polarity
configuration. The latter case simply corresponds to a flux rope configuration bounded below by
an X-type structure, i.e., in general a hyperbolic flux tube \cite[HFT; e.g.,][]{titov02}. It is
well known that this configuration admits of both stable and unstable states for the flux rope
(while the arcade below does not erupt): the helical kink mode
and the torus instability have been demonstrated to occur if the flux rope's twist or height exceed
a threshold \cite[e.g.,][]{tkt04,tk07}. The consideration of the former case in Paper~II
shows that it likewise admits of both stable and unstable states for the upper flux rope,
with the lower flux rope being stable (in addition to states with both ropes being unstable).
Consequently, both configurations sketched in \fig{cartoon} represent plausible models for the
observed double-decker filament.

% \newpage 

\subsection{Formation Mechanism} \label{ss:formation_mechanism}

A question naturally arises as to how a double-decker configuration containing two flux
ropes in equilibrium can form. We provide two possible scenarios as follows.

\begin{enumerate}[a)]
\item The lower branch emerges from below the photosphere after the upper branch has formed above
the PIL. This is motivated by the recent {\it Hinode} observation which suggests that a flux rope
can emerge under a pre-existing filament \citep{okamoto08}. In this particular
observation, the upper and lower flux systems appeared to merge within a couple of hours right after
the emergence of the lower flux system \citep{okamoto09}. Such merging may be delayed or inhibited
if the upper flux system is elevated to substantial heights or if the ambient field between the flux
systems is very strong (see Paper~II for the latter case).

\item Both branches originally belong to a single flux rope or flux bundle and are separated later.
This is motivated by the ``partial eruption'' scenario proposed by \citet{gilbert01}, in which the
reconnection within a stretched flux rope splits it into two ropes with the same handedness.
However, it is unclear whether the two ropes can remain in equilibrium for an extended period of
time (a few days), since in previous simulations \citep[e.g.,][]{gf06} and observations
\citep[e.g.,][]{gilbert01, lag07, gal07, tripathi09, regnier11} the splitting often occurs during
the eruption, or, in some cases, shortly before the eruption \citep[e.g.,][]{contarino03, guo10}.
See Paper~II for a new simulati0.5on addressing this problem.
\end{enumerate}

We suggest that a double-decker configuration is not a unique occurrence, since it is possible
for either branch to be void of filament material so that only a single-branch filament is visible.
In particular, it is sometimes observed that a filament survives the eruption directly above it
\citep[e.g.,][]{pevtsov02, liu07, lag07, liu10}. This could be similar to the eruption studied here
except that the upper branch is not traced by filament material. As the upper branch rises,
reconnection occurs between the oppositely-directed legs of the overlying field
which recloses between the two branches, so that the lower branch is confined by the newly
reconnected field \citep[similar to the middle panel of Figure~3 in][]{gilbert01}.

\subsection{Flux Transfer} \label{ss:flux_transfer}

The transfer of material from the lower to the upper filament branch implies a corresponding
transfer of flux if the field in the filament has a dominantly horizontal direction, which is the
standard configuration at least for active-region filaments. Such transfer of flux through the
HFT between the filament branches is different from a conventional reconnection process at the
HFT (e.g., tether-cutting reconnection), which would exchange flux in both branches with the
ambient flux. Thus it must be due to upward Lorentz forces of the current-carrying flux in the
lower branch which enforce part of this flux to rise even through the HFT. Such a process is
conceivable if part of the flux in the lower branch is particularly stressed (sheared or twisted)
by appropriate photospheric changes in its footpoint areas. It is also natural to expect that
such flux rearrangement is accompanied by reconnection between the flux of the lower branch and
that of the upper branch . We conclude that the transfer of flux as indicated by the observed
mass transfer must involve a transfer of current from the lower to the upper filament branch.
This will be further considered in the modeling of the partial eruption in Paper~II.

\subsection{Eruption Mechanism} \label{ss:mechanism}

\subsubsection{Role of the Helical Kink Instability} \label{sss:hKI}

The pronounced writhing of the upper filament branch into a projected forward S shape in the
course of its rise is an indication for the occurrence of the helical kink instability.
However, since the filament had a low-lying, reverse S shape prior to the eruption, it must
have temporarily straightened out during the initial rise before it could adopt the high-arching,
forward S shape. Both a low-lying reverse S shaped structure and a high-arching forward S shaped
structure have negative (left-handed) writhe, while a straight loop (lying in a plane) has no
writhe \citep{Torok&al2010}. Therefore, the straightening implies a reduction of writhe,
excluding the helical kink instability as the trigger of the eruption. This is because the
helical kink which transforms twist into writhe is supposed to increases the writhe. The only
exception is an opposite sign of the initial twist and writhe, which would be a very unusual case
and is in no way supported by the data of the considered event, which all indicate left-handed
helicity.

As has also been noted in \citet{Torok&al2010}, the initial increase of the flux rope twist by
the transformation of writhe helicity supports the occurrence of the helical kink in the further
evolution. Part of the acquired twist would thus be transformed back into left-handed writhe.
Inferring the occurrence of the instability from an observed writhing (helical kinking) alone is
often not conclusive, however, since a writhing of the same sign is given to a rising flux rope
by the shear component of the ambient field \citep{if07}. This component is usually present and
acts quite efficiently \citep{ktt12}. In AR~11093 it must have been rather strong because
the main polarities were situated near the ends of the filament. In order to infer the helical
kink unambiguously, one needs to find a super-critical amount of twist (through observation of
highly twisted substructure, magnetic field extrapolation, or a parametric study as in
\citeauthor{ktt12} \citeyear{ktt12}) or features incompatible with a writhing
driven (nearly) exclusively by the shear field. The latter---approaching flux rope legs, apex
rotation considerably exceeding 90~degrees, or more than one helical turn---were justified in
\citet{ktt12}.

In the present case, we find an indication against a purely shear field-driven writhing from the
filament shape in the \sat{sdo} images in \fig{erupt}(d)--(e). The filament exhibits two strong and
localized bends in this phase, which, at the given oblique perspective, indicate an apex rotation
exceeding 90~degrees (with respect to the line connecting the footpoints) by a considerable
amount. Projection effects can easily produce such a bend on one side of a gently writhed loop
with $\lesssim90$~degrees apex rotation. However, the other leg would then appear quite straight.
Two strong and localized bends are seen from many perspectives if the apex rotation clearly
exceeds 90~degrees. Such a strong rotation is not expected to result from the shear field
mechanism alone because the shear field causes the flux rope legs to lean to the side in opposite
direction perpendicular to its own direction. If there were no further effect, the resulting
rotation would always stay below 90~degrees. In fact, there is an additional contribution to the
rotation from the relaxation of the twist in a rising force-free flux rope even if the twist is
insufficient to trigger the helical kink. \citet{ktt12} have found that this contribution
is about 40~degrees if the shear field is very small, and it should be smaller for larger shear
field because the shear field-driven rotation then has a share in reducing the twist. Hence, a
total rotation of $\approx\!130$~degrees or more represents an indication for the occurrence of
the helical kink instability. The two strong bends in Figure~\ref{erupt}(d)--(e) are consistent
with a rotation of this magnitude. The alignment of the filament section between the bends with
the solar-$y$ axis in \fig{erupt}(e) is suggestive of a rotation beyond the meridional plane,
since the height of the structure is initially still increasing if one goes southward from the
northern bend (compare the \sat{stereo} image in \fig{erupt}(h)). If this section of the filament
were lying in the meridional plane, then it would run slightly southeastwards in the plane of the
sky for \sat{sdo}. This suggests a rotation even slightly exceeding 135~degrees with respect to
the diagonal line between the footpoints of the filament, which is an indication that the helical
kink instability did occur in the main phase of the eruption.

\subsubsection{Flux Imbalance and Torus Instability} \label{ss:TI}

Which other mechanism could have triggered this event? We suggest that it was a loss of
equilibrium caused by flux imbalance or by the torus instability. Both mechanisms, which may
actually be closely related, are supported by the main characteristics of the gradual evolution
prior to the event's onset: the transfer of flux from the lower to the upper filament branch and
the slow rise of the upper branch (\fig{slit}).

Through parametric study of flux ropes in numerical models of erupting active regions, it has
been found that the amount of axial flux in the rope, relative to the total amount of flux in the
region, possesses a limiting value for the existence of stable equilbria \citep{Bobra&al2008,
Savcheva&vanBallegooijen2009, Su&al2011}. The limiting value appears to be rather small, in the
$\sim(10\mbox{--}20)$~percent range (although compare \citeauthor{Green&al2011}
\citeyear{Green&al2011}, who presented support for a higher value for a flux rope still in the
process of formation). The conjectured flux rope in the present event was lying rather high, thus
likely well developed, so that the given small limiting value appears to be relevant and a rather
modest amount of flux transfer to the upper branch may have led it to a point where no
neighboring equilibrium was available.

The torus instability \citep{kt06, tk07} sets in if a flux rope rises to a critical height
at which the overlying field declines with height at a sufficiently steep rate \citep{Liu2008,
aulanier10, oz10, fan10}. Thus the observed slow rise of the upper filament branch makes this
instability a potential trigger mechanism.

\subsubsection{Mass (Un-) Loading} \label{ss:mass}

The observations in Section~\ref{ss:rise&eruption} show that mass is transferred from the lower
to the upper branch of the filament. Mass ``loading'' in some form may often play a role in the
final evolution of filaments toward an eruption, since their darkness and thickness often
increase in this phase \cite[e.g.,][]{kga09, guo10}. Since this darkening is not yet fully
understood, other effects, listed in Section~\ref{ss:rise&eruption}, may be relevant in addition,
or alternatively. Mass loading is very suggestive as a mechanism that helps holding down
current-carrying flux, thus raising the amount of free magnetic energy that can be stored in the
configuration \citep{lff03}. A destabilizing influence of ``mass unloading'' may also be
conjectured from the observation that the internal motions in filaments tend to amplify prior to
eruption, which happened also in the event analyzed here. Moreover, these motions were
systematically directed from the middle toward the ends of the upper branch for an extended
period of time, at least $\approx\!15$~minutes prior to the onset of the fast rise
(Section~\ref{ss:rise&eruption}). However, a consideration of the typical gravitational and
magnetic energy densities in active regions leaves mass unloading at most the role of a ``final
drop'' in an equilibrium sequence approaching the point where the equilibrium is lost.
\citet{Forbes2000} estimates that the magnetic energy of an average active region
($B\sim100$~Gauss) exceeds the gravitational energy by three orders of magnitude, based on a
typical coronal density of $10^9$~cm$^{-3}$. For the denser filament material, the measured
values of the electron density vary greatly, ranging from $10^9$ to $10^{11}$ cm${}^{-3}$
\citep{labrosse10}, due to differences among the techniques that have been used, as well as to an
unknown filling factor, but the temperature is often better constrained  ($T\le 10^4$ K). Thus,
to maintain a local pressure balance, filaments must be about 100 times denser than their typical
coronal surroundings ($T\simeq 10^6$ K, $n\simeq 10^9$ cm${}^{-3}$). Still, the energy density of
the gravitation is at least one order of magnitude smaller than that of the magnetic field. This
renders gravity largely irrelevant for the energy storage in active regions. Thus, while mass
unloading of the upper filament branch may have played a minor role in the evolution toward the
eruption, a role in the actual driving process can nearly certainly be excluded.

\subsection{Transient HXR Sigmoid}

An important structure associated with the dynamic evolution of a twisted flux rope is a
sigmoidal current sheet under the rope, as revealed in various MHD simulations. The current sheet
may form at a bald-patch separatrix surface \citep[BPSS; e.g.,][]{td99, ml01, gf06jgr, archon09,
fan10}, at a hyperbolic flux tube \citep[HFT; e.g.,][]{tgn03, gtn03, ktt04}, or simply in a layer
of highly sheared field \citep{tk03, apd05}. In each of these cases, a reverse S shaped current
sheet is associated with a left-handed flux rope whose axis writhes into a forward S shape when
the rope rises \citep{gf06jgr,ktt04}. The dissipation process in this current sheet and the
resultant heating of plasma are suggested to be responsible for transient sigmoidal structures
that brighten in soft X-rays prior to or during coronal eruptions. Our observation of the reverse
S shaped HXR sigmoid underlying the left-handed kink is also consistent with these simulations.

As demonstrated in \fig{erupt}, the coronal HXR sigmoid only formed after the upper branch of the
filament had risen to relatively high altitudes above the surface. As the X-ray sigmoid is largely of
coronal origin (between $\sim\,$12--25 Mm; recall Section~\ref{ss:3d} and see also \fig{sigmoid}(a)), it
is likely associated with the flare current sheet formed at the HFT.
We therefore suggest that accelerated electrons trapped in this current sheet produced the HXR
sigmoid.

\section{Summary} \label{s:summary}

We investigate the pre-eruptive evolution and the partial eruption of a filament which is
composed of two branches separated in height, combining \sat{sdo}, \sat{stereo}, and \sat{rhessi}
data. This is complemented by MHD modeling in Paper~II. To our knowledge, such a double-decker
configuration is analyzed in detail for the first time. We summarize the major results as follows.

\begin{itemize}
\item With stereoscopic observations from \sat{sdo} and \sat{stereo-B}, we obtain the
three-dimensional height information of the two filament branches. They are separated
in height by about 13 Mm, and the vertical extension of the upper branch is about 10 Mm.

\item The strong writhing of the upper branch into a left-handed helical kink, unambiguously
determined by combining \sat{sdo} and \sat{stereo-A} observations, clearly indicates the
structure of a left-handed flux rope for this branch. Since the lower branch has the same
magnetic connections at its ends and transfers some of its flux into the upper branch in the
course of the pre-eruptive evolution, left-handed helicity is indicated also for the lower
branch, hence for the dextral filament as a whole.

\item This structure is compatible with two model configurations, a flux rope above a sheared
magnetic arcade with dips and a double flux rope equilibrium. In either case,
the filament material in each branch can be supported against gravity by upward concave field
lines, and a hyperbolic flux tube separates the branches. The first configuration is well known
to possess stable and unstable states of the flux rope, with the arcade remaining stable
\citep{td99}. Equilibria of the second type are analytically and numerically constructed in
Paper II. MHD simulations demonstrate that they possess stable as well as unstable states
with only the upper flux rope erupting (and also unstable states that lead to the ejection
of both flux ropes). \\ 

\item The pre-eruptive evolution of the filament is characterized by a slow rise of the upper
branch most likely driven by the transfer of current-carrying flux from the lower to the upper branch in a
sequence of partial merging episodes. Weak flux cancelation may also have contributed to the
rise. These properties suggest that the eruption was triggered by reaching a point of flux
imbalance between the upper branch and the ambient field \cite[e.g.,][]{Su&al2011} or the
threshold of the torus instability.

\item The initial straightening of the erupting upper filament branch from its original reverse
S-shape excludes the helical kink instability as trigger of the eruption, but it supports the
occurrence of the instability in the main phase of the eruption, which is indeed indicated by the
strong forward S-shape acquired in this phase.

\item The main acceleration of the erupting branch commences very close in time with the
impulsive phase of the associated M1-class flare. The eruption results in a reverse S-shaped HXR
sigmoid which is at least partly located in the gap between the two branches from about 12~Mm to
25~Mm above the surface. To our knowledge, this is the first time that a coronal sigmoid is
clearly observed in HXRs. We suggest that electrons accelerated in the vertical (flare) current sheet
under the rising filament branch produced the coronal HXR emission, in agreement with a previous
model for transient sigmoids \citep{ktt04}.

\end{itemize}

\acknowledgments The authors are grateful to the \sat{sdo}, \sat{stereo} and \sat{rhessi} teams
for the free access to the data and the development of the data analysis software. KSO H$\alpha$
data are provided through the Global H-alpha Network operated by NJIT.R.~Liu acknowledges the
Thousand Young Talents Program of China, NSFC grants 41131065 and 41121003, 973 key project
2011CB811403, CAS Key Research Program KZZD-EW-01-4, and the fundamental research funds for the
central universities WK2080000031. R.~Liu, C.L., and H.W. were supported by NASA grants
NNX08-AJ23G and NNX08-AQ90G, and by NSF grants ATM-0849453 and ATM-0819662. B.K.\ acknowledges
support by the DFG and the STFC. The contributions of T.T., V.S.T., R.~Lionello, and J.A.L. were
supported by NASA's HTP, LWS, and SR\&T programs, and by CISM (an NSF Science and Technology
Center). Computational resources were provided by NSF TACC in Austin and by NASA NAS at Ames
Research Center. H.W. acknowledges travel support by Key Laboratory of Solar Activity, National
Astronomical Observatories of Chinese Academy of Sciences, under grant KLSA201201.

%\bibliographystyle{apj}
%\bibliography{ribbon}

\end{document}